\documentclass[prl,twocolumn,showpacs,groupedaddress]{revtex4}

\usepackage{graphicx}
\usepackage{dcolumn}
\usepackage{bm}
\usepackage{hyperref}
\usepackage{amssymb}
\usepackage{amsmath}
\usepackage{epsfig}

\begin{document}

\title{The final mass and spin of black hole mergers}

\author{Wolfgang Tichy}
\affiliation{Department of Physics,
	Florida Atlantic University, 
	Boca Raton, FL 33431, USA}
\author{Pedro Marronetti}
\affiliation{Department of Physics,
	Florida Atlantic University, 
	Boca Raton, FL 33431, USA}
	
\pacs{
04.25.D-,	
04.25.dg,	
04.70.-s,	
97.60.Lf,	
98.62.Js,	
98.65.Fz 	
}

%
\newcommand\ba{\begin{eqnarray}}
\newcommand\ea{\end{eqnarray}}
\newcommand\et{{\it et al.~}}
\newcommand{\rem}[2]{{\bf #1}:\\ {\it #2}\\}
%

\begin{abstract}

We consider black holes resulting 
from binary black hole mergers. By fitting to numerical results
we construct analytic formulas that predict the
mass and spin of the final black hole. Our formulas are
valid for arbitrary initial spins and mass ratios and agree 
well with available numerical simulations.
We use our spin formula in the context of two common merger scenarios
for supermassive galactic black holes.
We consider the case of isotropically distributed initial spin 
orientations (when no surrounding matter is present)
and also the case when matter closely aligns the spins with the
orbital angular momentum.
The spin magnitude of black holes resulting from
successive generations of mergers (with symmetric mass
ratio $\eta$) has a mean of $1.73\eta + 0.28$ in the isotropic case 
and $0.94$ for the closely aligned case.

\end{abstract}

\maketitle


{\bf Introduction.} 
Barring the influence of surrounding matter or third objects, 
two gravitationally bound black holes (BHs)
(with masses $M_a$, $M_b$ and spins $\vec{S_a}$,$\vec{S_b}$)
will orbit around their common center of mass emitting gravitational
radiation, which carries away energy, momentum and angular momentum.
This radiation will circularize the orbits of
the progenitor BHs and eventually shrink the orbits until the BHs
merge and form a BH of mass $M_f$ and spin $S_f$.
Thus, the initial state is described by eight parameters:
the mass ratio $q \equiv ~M_b/M_a$, the dimensionless
spins $\vec{a}=\vec{S_a}/M_a^2$, $\vec{b}=\vec{S_b}/M_b^2$ 
of the initial BHs and the
dimensionless angular velocity $\omega = \Omega (M_a+M_b)$. 
Here $\omega$ specifies which point on the possibly very long 
inspiral trajectory is used as initial point.
After the merger the final BH is characterized by seven
parameters, the final mass $m=M_f/(M_a+M_b)$, the
spin $\vec{s}=\vec{S_f}/M_f^2$ and the kick velocity $\vec{k}$.
Predicting the final $m$ and $\vec{s}$ from the initial 
parameters is of great importance in many astrophysical
merger scenarios.
Boyle, Kesden and Nissanke~\cite{Boyle:2007sz,Boyle:2007ru} propose to
describe any of the final parameters as a Taylor expansion in the six initial
spin components. These spin expansions (with coefficients 
fitted to date from numerical simulations)
are based on the assumption that any dimensionless final quantity 
must be a function of the eight initial parameters.
In this paper we use their expansions for the final dimensionless 
mass and spin. Note, however, that all expansion coefficients
still depend on $q$ and $\omega$, since we Taylor expand
only in $a_i$ and $b_i$. I.e. even if we know the
coefficients for a particular mass ratio and a particular
initial angular velocity we cannot predict any final quantities 
for different mass ratios or initial velocities.
At first glance this seems to severely limit the usefulness
of these expansions. However, as we will see, we have come
up with a particular fit for the $q$-dependence, which seems
to work rather well. Also, we will show that the final spin
magnitude depends only weakly on $\omega$. The
orientation of the final spin, however, does depend on $\omega$
as one would expect due to spin precession of the individual
BH spins if one starts at a different $\omega$.
Nevertheless, our approach can even give approximate spin orientations
if we start with values for $\omega$ like the ones used here,
which are typical for the current state-of-the-art simulations performed
by most groups.

We use our formulas to calculate the probability density of the
magnitude of the final spin for successive generations of
BH mergers. We consider ``gas-rich" or
``wet" mergers, where a circumbinary disk surrounds the binary,
and also ``gas-poor" or ``dry" mergers, where no matter is 
present~\cite{Bogdanovic:2007hp,Berti:2008af}.


{\bf Spin expansions.} Our coordinate system is such that
the $z$ axis is perpendicular to the initial orbital plane. 
The center of mass is initially at rest at the origin, with the
$x$ axis along direction of the momentum of BH b. 
The $y$ axis is along the line connecting the BHs
with BH a located at $y>0$.

To construct formulas aimed at predicting the final BH 
mass and spin, we follow the method described in~\cite{Boyle:2007ru}. 
To linear order,
the spin expansions for the final mass and spin are:
\ba
m &=& m^0 + (m^{a1} a_z + m^{b1} b_z) \nonumber \\ 
s_x &=& (s_x^{a1} a_x + s_x^{b1} b_x) + (s_x^{a2} a_y + s_x^{b2} b_y) \nonumber\\
s_y &=& (s_y^{a1} a_x + s_y^{b1} b_x) + (s_y^{a2} a_y + s_y^{b2} b_y) \nonumber\\
s_z &=& s_z^0 + (s_z^{a1} a_z + s_z^{b1} b_z) ,
\label{spin_exp}
\ea
where the coefficients are functions of the mass ratio $q$ and the 
initial orbital angular velocity $\omega$.
Notice that by using symmetries such as parity or exchange
many terms that would appear
in an unconstrained Taylor expansion have been dropped.
In addition, all the coefficients enclosed in common brackets are related,
by $m^{b1}(q,\omega) = m^{a1}(1/q,\omega)$ and
$s_i^{bj}(q,\omega) = s_i^{aj}(1/q,\omega)$.

We ignore the dependence on $\omega$ for the
time being and focus first on the equal mass case.
Note that for $q=1$ the above mentioned relation between coefficients
implies that all terms enclosed in common brackets have equal coefficients.
In order to determine these coefficients we have
performed 10 numerical simulations of equal mass binaries
with spins of magnitudes between 0.1 and 0.27 with
orientations as in Table XI of~\cite{Boyle:2007ru}.
All 10 simulations start with $\omega=0.05$ and are performed using 
the ``moving punctures'' method~\cite{Campanelli:2005dd,Baker:2005vv}
with the BAM code~\cite{Bruegmann:2003aw,Bruegmann:2006at}
which allows us to use moving nested refinement boxes.
We use 10 levels of 2:1 refinements. The outer boundaries
are located $436M$ away from the initial center of mass
($M$ being the sum of the initial BH masses),
and our resolution ranges between $8M$ on the outermost box
to $M/64$ near the BHs. 
Apart from the use of sixth order stencils in the interior of 
the boxes~\cite{Husa:2007hp}, our setup and methods to
determine spin and mass are very similar to the simulations 
reported in~\cite{Marronetti:2007ya,Tichy:2007hk,Marronetti:2007wz}.
Our 10 simulations are in principle sufficient to determine
all coefficients appearing in $s_i$ and $m$ up to quadratic order.
However, since the numerical errors in determining the final 
mass and spin are 0.1\% and 0.5\% respectively,
and since all quadratic coefficients are small, the errors in
all quadratic coefficients are more than 100\%. For this reason
only the linear coefficients are listed in Table \ref{lin_coeffs}.
\begin{table}
{\small
\begin{tabular}{l}
 $m^0 = 0.9515\pm 0.001$ \       $m^{a1}=m^{b1}=-0.013\pm 0.007$ \\
$s_x^{a1}=s_x^{b1} = 0.187\pm 0.002 =+s_y^{a2}=+s_x^{b2}$ \\
$s_y^{a1}=s_x^{b1} = 0.028\pm 0.002 =-s_x^{a2}=-s_x^{b2}$ \\
$s_z^0 = 0.686\pm 0.004$ \ \ \   $s_z^{a1}=s_z^{b1}=0.15\pm 0.03$ \\
\end{tabular}
}
\caption{\label{lin_coeffs}
Equal mass coefficients up to linear order.
}
\end{table}
In order to verify that linear expansions with the coefficients
in Table~\ref{lin_coeffs} give reasonable results
we have performed 10 more equal mass runs with spins of magnitude
0.75 with arbitrary orientations. As one can see from the
first 10 lines in Table~\ref{newruns_table}, the agreement between
our additional runs and the values predicted by the expansions
is quite good.
\begin{table}
{\small
\begin{tabular}{r|r|r|r|r|r|r|r|r|r|r}
$q$\ &$a_x$\ &$a_y$\ &$a_z$\ &$b_x$\ &$b_y$\ &$b_z$\ &$s$\ &$s_p$\ &$m$\ &$m_p$\ \\
\hline
1:1  &  .750 &  .000 & -.015 &  .750 &  .000 & -.015 &.772 & .755  &.949 & .952\\
1:1  &  .739 & -.128 &  .000 &  .095 & -.680 &  .303 &.777 & .773  &.945 & .948\\
1:1  & -.721 & -.061 &  .199 &  .280 & -.159 & -.677 &.622 & .621  &.956 & .958\\
1:1  & -.146 & -.704 & -.215 & -.413 & -.333 &  .530 &.778 & .776  &.943 & .947\\
1:1  & -.639 & -.227 &  .321 & -.517 & -.543 &  .020 &.795 & .796  &.939 & .947\\
1:1  & -.431 &  .340 & -.511 &  .187 & -.588 & -.426 &.542 & .544  &.960 & .964\\
1:1  & -.443 &  .323 &  .512 &  .390 &  .324 &  .553 &.849 & .849  &.929 & .938\\
1:1  &  .136 &  .405 & -.616 &  .082 & -.704 &  .244 &.635 & .634  &.955 & .956\\
1:1  &  .010 &  .187 &  .726 &  .174 & -.489 &  .542 &.867 & .867  &.925 & .935\\
1:1  &  .006 &  .054 & -.748 &  .398 & -.635 & -.026 &.587 & .585  &.958 & .962\\
5:6  &  .000 &  .000 &  .000 &  .000 &  .000 &  .000 &.682 & .682 &.952  & .952\\
5:6  &  .200 &  .000 & -.001 &  .000 &  .000 &  .000 &.684 & .684 &.952  & .952\\
5:6  &  .000 &  .150 & -.000 &  .000 &  .000 &  .000 &.683 & .683 &.952  & .952\\
5:6  &  .000 &  .000 &  .000 &  .200 &  .000 & -.000 &.683 & .683 &.952  & .952\\
5:6  &  .000 &  .000 &  .000 &  .000 &  .150 & -.000 &.682 & .682 &.952  & .952\\
5:6  &  .200 &  .000 &  .099 &  .000 &  .000 &  .000 &.703 & .702 &.950  & .951\\
5:6  &  .000 &  .150 & -.100 &  .000 &  .000 &  .000 &.664 & .664 &.953  & .953\\
5:6  &  .000 &  .000 &  .000 &  .200 &  .000 &  .100 &.695 & .695 &.951  & .951\\
5:6  &  .000 &  .000 &  .000 &  .000 &  .150 & -.100 &.671 & .670 &.953  & .953\\
5:6  &  .000 &  .150 & -.101 &  .000 &  .150 & -.001 &.667 & .666 &.953  & .953\\
5:6  &  .200 &  .000 & -.001 & -.000 &  .150 & -.100 &.673 & .672 &.953  & .953\\
5:6  &  .000 &  .150 & -.000 &  .200 &  .000 &  .100 &.696 & .696 &.951  & .951\\
5:6  &  .200 &  .150 & -.001 &  .000 &  .000 &  .000 &.685 & .685 &.952  & .952\\
5:6  &  .000 &  .000 &  .000 &  .200 &  .150 & -.001 &.683 & .683 &.952  & .952\\
5:6  &  .000 &  .000 &  .100 &  .000 &  .000 &  .100 &.712 & .712 &.949  & .949\\
5:8  &  .091 &  .268 & -.412 &  .043 & -.375 &  .132 &.565 & .567  &.961 & .957\\
5:8  & -.287 &  .225 & -.342 &  .100 & -.315 & -.226 &.556 & .558  &.962 & .961\\
2:3  & -.384 & -.033 &  .106 &  .112 & -.064 & -.271 &.676 & .674  &.955 & .955\\
2:3  & -.339 & -.120 &  .174 & -.207 & -.217 &  .011 &.724 & .723  &.950 & .951\\
2:3  & -.230 &  .180 & -.273 &  .075 & -.236 & -.169 &.587 & .588  &.960 & .959\\
2:3  &  .073 &  .214 & -.330 &  .033 & -.281 &  .099 &.596 & .597  &.959 & .956\\
\end{tabular}
}
\caption{\label{newruns_table}
Test simulations: the columns
show the initial mass ratio $q$, the progenitor spins $a_i$ and $b_i$, 
the final spin magnitude $s$ and mass $m$, and 
our predictions $s_p$ and $m_p$.
}
\end{table}
This demonstrates that for $q=1$ we can trust the expansions 
with our coefficients. 
As we can see from Table \ref{lin_coeffs}, $s_x^{a1}=s_y^{a2}$
and $ s_x^{a2}=-s_y^{a1} $, which is not predicted
by the symmetries used to derive the spin expansions.
Thus, in terms of the initial spins $\vec{S}_{a}$, $\vec{S}_{b}$ and 
final spin $\vec{S}_{f}$ the $x$ and $y$
components of Eq.~(\ref{spin_exp}) can be rewritten as
\begin{equation}
\begin{pmatrix}
\label{Sxyfinal}
 S_{f,x}\\
 S_{f,y}
\end{pmatrix}
= \alpha R 
\begin{pmatrix}
 S_{a,x} + S_{b,x} \\
 S_{a,y} + S_{b,y}
\end{pmatrix}
\end{equation}
with
\begin{equation}
\label{RotScal}
R =
\begin{pmatrix}
 \cos\beta      & -\sin\beta \\
 \sin\beta      &  \cos\beta
\end{pmatrix}, \ 
\beta=0.15, \
\alpha = 0.685.
\end{equation}
This means that the final spin components in the $xy$-plane
are given by rotating and scaling the initial spins. The
scale factor is $\alpha = 0.685$ and the rotation matrix is $R$.
This result is physically reasonable, because we know that
the initial spins will precess and also radiate some angular momentum.
If Eq.~(\ref{Sxyfinal}) also held for unequal
masses it would imply a certain form for the mass ratio dependence:
\begin{equation}
\begin{pmatrix}
 s_x\\
 s_y
\end{pmatrix}
=
\alpha R 
\begin{pmatrix}
 \frac{M_a^2}{M_f^2} a_x + \frac{M_b^2}{M_f^2} b_x \\
 \frac{M_a^2}{M_f^2} a_y + \frac{M_b^2}{M_f^2} b_y
\end{pmatrix}
\approx
\alpha R
\begin{pmatrix}
 \frac{a_x}{(1+q)^2} + \frac{q^2 b_x}{(1+q)^2} \\
 \frac{a_y}{(1+q)^2} + \frac{q^2 b_y}{(1+q)^2}
\end{pmatrix}
\end{equation}
However, the latter cannot be true for extreme mass ratios, since
than we would not get $\vec{s} = \vec{a}$ for $q=0$. In order to
get the extreme mass ratio limit correctly, we modify
the mass ratio dependence somewhat and assume that
\begin{equation}
\label{sxyfinal}
\begin{pmatrix}
 s_x\\
 s_y
\end{pmatrix}
=
\begin{pmatrix}
 s_x^{a1} \left[ g(q,c) a_x + g(1/q,c) b_x \right]
-s_y^{a1} 4\eta \left(a_y + b_y\right)
\\
 s_y^{a1} 4\eta \left(a_x + b_x\right)
+s_x^{a1} \left[ g(q,c) a_y + g(1/q,c) b_y \right]
\end{pmatrix},
\end{equation}
where 
\begin{equation}
\label{sxycoeffs}
\eta = \frac{q}{(1+q)^2}, \ 
g(q,c) = \frac{(c+1)^2}{(c+q)^2}, \ 
c    = \frac{\sqrt{s_x^{a1}}}{1-\sqrt{s_x^{a1}}}=0.762,
\end{equation}
are chosen such that $\vec{s} = \vec{a}$ for $q=0$ 
and $\vec{s} = \vec{b}$ for $q=\infty$.

So far we have only considered the components in $xy$-plane.
For equal masses $s_z^{a1}=s_z^{b1}$ so that $s_z$ in Eq.~(\ref{spin_exp})
simplifies. As above we introduce a $q$-dependence given by
\begin{equation}
s_z = s_z^{0}( 4 w \eta + 16(1-w) \eta^2) + 
  s_z^{a1} \left[ g(q,c_3) a_z + g(1/q,c_3) b_z  \right],
\end{equation}
where $c_3  = \sqrt{s_z^{a1}}/(1-\sqrt{s_z^{a1}})=0.632$.
The $\eta$ dependence of the leading coefficient is inspired
by post-Newtonian (PN) expressions, and has been shown to fit runs
without initial spins very well~\cite{Berti:2007fi}.
The constant $w = 1.26$ is fitted using the results for
unequal masses discussed below.

Similarly, we build a mass ratio dependence into the final mass
formula and write
\begin{equation}
\label{mfinal}
m = 1 + (m^{0}-1) 4\eta + m^{a1} 16\eta^2 \left( a_z + b_z  \right).
\end{equation}
The leading term is inspired by the fact that the binding
energy is proportional to $\eta$, and the linear term is chosen
such that it has the correct limit in the extreme mass ratio case.

While the $q$ dependence we have introduced so far reproduces
our fits for the equal mass case and simultaneously gives the
correct answers in the extreme mass ratio cases, it is not clear
how well our formulas perform for intermediate mass ratios.
For this reason we have performed 21 more simulations with $q$ different
from unity. As one can see from Table~\ref{newruns_table}
the predictions are at most 0.4\% from the numerical results.

So far we have only kept linear terms in our expansions. We find that
our formulas agree with a wide range of test runs. Yet, for initial
spin magnitudes close to 1 our formulas deviate from the extrapolated
values~\cite{Campanelli:2006fy,Marronetti:2007wz,Rezzolla:2007xa}
for the minimum and maximum possible final spin.
These problems can be fixed if we add some quadratic terms to $s_z$ alone:
\begin{eqnarray}
\label{szfinal}
s_z &=& s_z^{(0)}( 4 w \eta + 16(1-w) \eta^2) \nonumber \\
&& +  s_z^{a1} \left[ g(q,c_3) a_z + g(1/q,c_3) b_z  \right] \nonumber \\
&& + 16 k \eta^2 [(a_x+b_x)^2 +(a_y+b_y)^2 -(a_z+b_z)^2].
\end{eqnarray}
These quadratic terms approximately reproduce most of the 
very small and very uncertain quadratic coefficients in $s_z$.
The coefficient $k=0.008$ is chosen such that we get the
best overall agreement with our numerical simulations.


{\bf Results.} Our particular mapping of initial masses and
spins into the final mass and spin given by 
Eqs.~(\ref{mfinal}), (\ref{sxyfinal}), (\ref{sxycoeffs}) and (\ref{szfinal})
was fitted for the initial orbital
angular velocity $\omega = 0.05$. If we start with
the same initial spin components but at lower initial angular velocity
(i.e. larger separation) the individual spins and the orbital
plane will have precessed by the time we reach $\omega = 0.05$.
In this case we cannot expect that our formulas
will predict the final spin components if we simply use
the initial spin components. However, the final spin magnitude
should still be approximately correct, since PN 
calculations~\cite{Kidder:1995zr} 
demonstrate that the spin magnitudes are conserved at 2PN order.
This expectation is borne out for the following additional $q=1$
test run.
It starts with $\omega = 0.03$,
$\vec{a} = (-0.637,-0.226,0.325)$,
$\vec{b} = (-0.517,-0.543,0.025)$,
and yields a final spin
$\vec{s} = (-0.226,-0.146,0.746)$
with magnitude
$s = 0.793$.
Our fitting formula predicts
$\vec{s} = (-0.194,-0.176,0.753)$
with magnitude
$s = 0.797$.
This shows that the predicted magnitude is correct up to an error
of 0.6\%, while the components only agree if the predicted vector
is rotated. 
This rotation comes from the precession during the time
it takes to go from $\omega = 0.03$ to $\omega = 0.05$, during which
the system completes about 4 orbits.

We have also compared with the numerical results published
in~\cite{Berti:2007fi,Tichy:2007hk,Marronetti:2007wz,
Rezzolla:2007rz,Rezzolla:2007xa,Dain:2008ck,Lousto:2008dn,Baker:2008mj},
which all start from an $\omega$ that is not too far from $0.05$.
The average deviation of all these results from our predictions
is about 1\%.
Hence our formulas gives useful predictions for the final
mass and even for the spin components if one starts
near $\omega = 0.05$.
However, if one is ever interested in the final spin
components for a much larger initial separation with $\omega \ll 0.05$,
one can still evolve the spins using PN theory up to
$\omega = 0.05$ and then use our formula. This eliminates the need
for expensive numerical simulations.

Other groups have presented formulas that, like ours, attempt to predict the
final spin of the merger.
The analytic estimate of Buonanno et al.~\cite{Buonanno:2007sv}
can give the final spin magnitude to within few percent with
larger deviations for spins close to anti alignment.
Starting from a number of assumptions, Rezzolla et al.~\cite{Rezzolla:2007rz}
developed a more accurate formula with coefficients
fitted to numerical results. One of their assumptions
is that the components of the final spin in
the initial orbital plane are obtained by summing the 
components of the initial spins i.e., $S_{f,x/y}= S_{a,x/y}+ S_{b,x/y}$. 
In our simulations we observe that this assumption is violated.
From the discussion around Eqs.~\ref{Sxyfinal} and \ref{RotScal}
it is clear that these components get slightly rotated and
shortened by a factor $\alpha=0.685$ during the merger.
For instance, the run from the first line in Table \ref{newruns_table}
we find a value of $s_{xy} = 0.291$ for the in-plane component of the spin.
Our formula (\ref{sxyfinal}) predicts a close value of $s_{xy,p} = 0.284$,
the approach in~\cite{Rezzolla:2007rz} leads to $s_{xy,R} = 0.375$
which is about 30\% too large, while both approaches predict almost
the same spin magnitude. Thus our formulas improve the final spin
orientation.

\begin{figure}
\includegraphics[scale=0.3,clip=true]{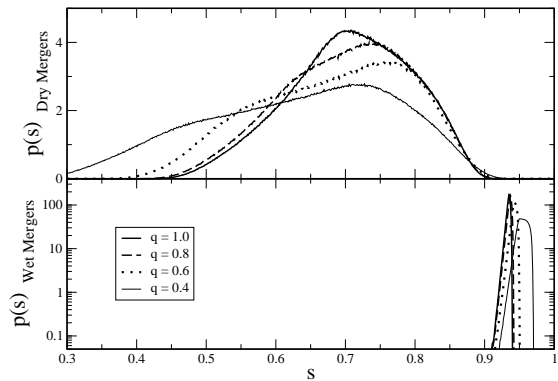}
\caption{\label{Fig_Last_Gen}
Probability density (for different mass ratios) 
of the final BH spin magnitude after
4 generations of ``dry" mergers (top) and ``wet" mergers (bottom).
}
\end{figure}
\begin{figure}
\includegraphics[scale=0.3,clip=true]{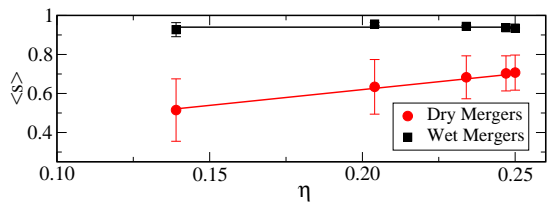}
\caption{\label{Fig_Mean_jm2}
Mean value of the final spin magnitude as a function of the symmetric 
mass ratio $\eta$. The vertical bars are the corresponding standard
deviations.
}
\end{figure}

{\bf Black hole mergers.} Using Eqs.~(\ref{sxyfinal}), (\ref{sxycoeffs}) 
and (\ref{szfinal}) we can study the properties
of the spin of BHs produced by successive binary BH mergers.
Two types of scenarios are likely to witness mergers: one
in which the two BHs carry with them dense surrounding matter
(``wet" mergers) and another where the progenitors meet in relatively empty
space (``dry" mergers)~\cite{Bogdanovic:2007hp}. In the former case the
progenitors spins are likely to reach the merger aligned with the orbital
angular momentum~\cite{Schnittman:2004vq}, while in the latter the same
spins are bound to be isotropically oriented~\cite{Bogdanovic:2007hp}. In
our models we assume that the initial spin directions either have a uniform
probability distribution (``dry" mergers) or that the longitudinal angle
$\theta$ obeys  a normal distribution centered in the direction of the
orbital angular momentum (``wet" mergers) 
\footnote{We use a standard deviation of $\sigma=\pi/20$, but almost
identical results are obtained for $\pi/40 \le \sigma \le \pi/5$.}.
For the spin magnitude we assume
that the original progenitors' spin magnitude is uniformly
distributed in the interval $[0,1]$.
The probability density for the final spin magnitude after
merger can then be obtained with Monte Carlo simulations. 
Note that the $\omega$ dependence of our formula is removed by
the integration over the spin orientation. This 
procedure can be iterated for successive generations assuming that
the magnitude of the progenitors spins has the probability density of the
previous generation. The resulting probability distribution
converges quickly to the curves shown in Fig.~\ref{Fig_Last_Gen} 
for the mass ratios $q=0.4,~0.6,~0.8$ and $1.0$.
The mean values for the probability densities as a function of the symmetric
mass ratio $\eta$ are shown in Fig.~\ref{Fig_Mean_jm2} with the vertical
bars corresponding to the standard deviations. The straight lines are linear
fits of the data points.
We find that successive ``dry" mergers produce final
spins with a mean value of
$\langle s\rangle_{Dry} = 1.73 \eta + 0.28$.
For $q=1$ the final spin does not
exceed $s=0.954$ (as our formula predicts for maximal aligned spins).
Successive ``wet" mergers produce a spin around 0.94, with
very little spread. 
Similar results have also been found by Berti et al.~\cite{Berti:2008af}
using the spin formula in~\cite{Rezzolla:2007rz}. This agreement is
expected since our spin formulas give magnitudes in close
agreement with ~\cite{Rezzolla:2007rz}. The main difference
is in the spin orientation which has been integrated out.

{\bf Discussion.} 
By fitting to numerical results we construct formulas 
[Eqs.~(\ref{mfinal}), (\ref{sxyfinal}), (\ref{szfinal})]
that predict the mass and spin of the final BH coming from binary
BH mergers.
We use them to determine the probability distribution
of the final spin magnitude (Figs.~\ref{Fig_Last_Gen},~\ref{Fig_Mean_jm2})
after several generations of mergers of either ``dry" or ``wet" mergers.

\begin{acknowledgments}

It is a pleasure to thank B. Br\"ugmann, L. Rezzolla, E. Barausse,
L. Boyle and M. Kesden for useful discussions.
This work was supported by NSF grant PHY-0652874. 
We acknowledge TACC at UT Austin for providing HPC resources
under allocations TG-PHY080022N and TG-MCA08X010.

\end{acknowledgments}

\bibliography{references}

\end{document}